\documentclass[
preprint,
 amsmath,amssymb,
 aps,
prb,
floatfix,
]{revtex4-2}

\usepackage{graphicx}
\usepackage{dcolumn}
\usepackage{booktabs, tabularx, ragged2e}
\usepackage{bm}
\usepackage[mathlines]{lineno}
\usepackage{commath}
\usepackage{physics}
\usepackage{color}
\usepackage{mathtools}

\begin{document}

\preprint{APS/123-QED}

\title{One-dimensional confined Rashba states in a two-dimensional Si$_{2}$Bi$_{2}$ induced by vacancy line defects}

\author{Arif Lukmantoro}
  \affiliation{
 Departement of Physics, Faculty of Mathematics and Natural Sciences, Universitas Gadjah Mada, Sekip Utara BLS 21 Yogyakarta 55186 Indonesia}
 \affiliation{
Research Center for Quantum Physics, National Research and Innovation Agency, Tangerang Selatan 15314, Indonesia.
}

\author{Edi Suprayoga}
\affiliation{
Research Center for Quantum Physics, National Research and Innovation Agency, Tangerang Selatan 15314, Indonesia.
}
 
\author{Moh.~Adhib~Ulil~Absor}%
 \email{adib@ugm.ac.id}
  \affiliation{
 Departement of Physics, Faculty of Mathematics and Natural Sciences, Universitas Gadjah Mada, Sekip Utara BLS 21 Yogyakarta 55186 Indonesia}

\date{\today}

\begin{abstract}
Advanced defect engineering techniques have enabled the creation of unique quantum phases from pristine materials. One-dimensional (1D) atomic defects in low-dimensional systems are particularly intriguing due to their distinct quantum properties, such as 1D Rashba states that allow for the generation of nondissipative spin currents, making them ideal for spintronic devices. Using density-functional calculations and model-based symmetry analysis, we report the emergence of 1D Rashba states in a two-dimensional Si$_{2}$Bi$_{2}$ monolayer (ML) with vacancy line defects (VLDs). We show that introducing VLDs in the Si$_{2}$Bi$_{2}$ ML induces 1D confined defect states near the Fermi level, which are strongly localized along the extended defect line. Notably, we observed 1D Rashba spin-split bands in these defect states with significant spin splitting originating mainly from the strong $p-p$ coupling orbitals between Si and Bi atoms near the defect sites. These spin-split defect states exhibit perfectly collinear spin polarization in momentum $\vec{k}$-space, which is oriented perpendicularly to the VLD orientation. Moreover, using $\vec{k}\cdot\vec{p}$ perturbation theory supplemented with symmetry analysis, we show that the 1D Rashba states with collinear spin polarization are enforced by the lowering of symmetry of the VLDs into the $C_{s}$ point group, which retains the $M_{xz}$ mirror symmetry along with the 1D nature of the VLDs. The observed 1D Rashba states in this system protect carriers against spin decoherence and support an exceptionally long spin lifetime, which could be promising for developing highly efficient spintronic devices.
\end{abstract}

\pacs{Valid PACS appear here}
\maketitle

\section{INTRODUCTION}

The successful synthesis of atomically precise low dimensional materials has led to a variety of intriguing quantum phases on the mesoscopic scale. These include the proximity of $p$-wave superconductivity and charge density wave \cite{Xi2016, Wickramaratne}, the topological Weyl semimetallic properties with significant magnetoresistance \cite{Ali2014, Li2017}, as well as the unique orbital and quantum spin Hall effects \cite{Xiaofeng, Bhowal}. One-dimensional (1D) systems, where electrons are confined to a 1D structure, have become particularly fascinating due to their unique quantum characteristics and potential applications in electronic devices. For example, in a 1D system, electron-phonon interactions can lead to the Peierls phase transition, a metal-insulator transition caused by charge density waves \cite{Yeom}. Additionally, 1D dimerization of the atomic lattice can create a topologically protected gap along the wire, with gapless solitonic excitations observed at the edges \cite{Pernet2022}. Recent reports indicate that electron correlations in 1D metallic systems can produce exotic quantum states characteristic of Tomonaga-Luttinger liquid (TLL) behavior \cite{Jia2022}. Coupled with strong spin-orbit coupling (SOC), these 1D metallic systems are also predicted to host zero-energy states known as Majorana bound states, enabling spinless $p$-wave Cooper pairing through a semiconductor nanowire with a helical state and inducing $s$-wave superconductivity \cite{Chen2020, ZhangXW, Zhang2021}. These significant properties, along with the topological protection afforded by the bulk-surface correspondence in topologically nontrivial quantum materials, making 1D electronic systems as promising candidates for next-generation electronic devices.

Another phenomenon of interest in 1D systems is the emergence of the 1D Rashba effect. The Rashba effect refers to the spin-splitting in electronic bands due to the SOC that appears in the case of broken structural inversion symmetry in a crystal due to the presence of a surface or an interface \cite{Rashba}. The Rashba effect has gained renewed interest following the discovery of topological insulators due to its significant potential in spintronics \cite{Frantzeskakis, Lyu2018}, where Rashba SOC is vital for creating and controlling spin currents \cite{Manchon, MihaiMiron2010, Sanchez2013}. Although the Rashba effect is commonly associated with two-dimensional (2D) systems, including surfaces \cite{Varykhalov, Go, Usachov}, heterointerface \cite{Geldiyev, Kong}, and two-dimensional layered structures \cite{WuKai, Yao2017, Absor2023, Adhib, Lukmantoro}, it has also been experimentally realized in 1D systems, such as gold (Au) chains on the vicinal silicon (Si) (111) \cite{Barke} and Si (557) surfaces \cite{Okuda}, platinum (Pt) nanowires on a Si (110) surface \cite{Park}, lead (Pb) nanoribbons on a Si (553) surface \cite{Kopciuszynski2017}, and the edges of Bi island on a Si(111) surface \cite{Takayam}. In addition, the 1D Rashba states have been theoretically predicted on  Pb atomic 1D chains on a semiconductor surface \cite{Mihalyuk}, Gd-adsorbed 1D zigzag graphene nanoribbon \cite{Qin}, and 1D zigzag bismuth nanoribbons \cite{Naumov}. Compared to 2D systems, 1D Rashba materials are favored because the potential gradients in 1D can be larger due to lower symmetry \cite{Tanaka}, and hence the SOC effects are significantly enhanced in reduced dimensions \cite{Soumyanarayanan2016}. Additionally, the unique unidimensional dispersion of the Fermi surface in 1D Rashba systems with unidirectional spin polarization allows for the generation of nondissipative spin currents \cite{Kammermeier, Schliemann}, making them ideal for spintronic devices.

Although 1D Rashba materials have been successfully fabricated using pure heavy metal chains artificially grown on a silicon surface \cite{Barke, Okuda, Park, Kopciuszynski2017}, this method presents significant practical difficulties. The main issue is the requirement for an in situ growth environment and precise control over various parameters \cite{Guo2022}. An alternative method for achieving the 1D Rashba system involves introducing 1D atomic line defects, such as vacancy line defects \cite{Chen2020, Chen2022, Zhang2021}, grain boundaries \cite{Zhou2024, Mesaros2024}, and edges \cite{Peng2017}, which has been developed by using advanced techniques like atomic force microscopy, scanning tunneling microscopy, and electron beam lithography. In particular, we focus on 2D materials with atomic thickness and high surface atom exposure, enabling easy property regulation through defect engineering \cite{Santra2024, Jie}. However, to the best of our knowledge, only a few classes of 2D materials with 1D atomic defects have been reported to support 1D Rashba states, including Fe(Te)Se monolayer (ML) \cite{Chen2020, Zhang2021, Wu2023} and several 2D transition metal dichalcogenides (TMDCs) such as $1H$-$MX_{2}$ ($M$: W, Mo; $X$: S, Se) MLs \cite{ZhangXW, Li2019}, $1T$-PtSe${2}$ ML \cite{Adhib}, and $1T'$ WTe${2}$ ML \cite{Absor2024}. This motivates the search for 1D defective systems in 2D materials that exhibit the 1D Rashba effect, as it has the potential to broaden the range of materials suitable for spintronics applications.

In this paper, through systematic study using first-principles density-functional theory (DFT) calculations and model-based symmetry analysis, we report the emergence of the 1D Rashba states in a 2D silicon bismotide (Si$_{2}$Bi$_{2}$) ML under the present of the vacancy line defects (VLDs). The Si$_{2}$Bi$_{2}$ ML belongs to the 2D group IV-V compounds, and its stability and electronic properties have been previously reported \cite{Ozdamar, Bafekry, Lee2020, Lukmantoro, Absor_Arif}. We reveal that introducing VLDs in the Si$_{2}$Bi$_{2}$ ML induces 1D confined defect states near the Fermi level, which are highly localized along the extended defect line. Notably, these defect states exhibit the significant 1D Rashba spin splitting, which is primarily determined by the strong $p-p$ coupling orbitals between the Si and Bi atoms near the defect sites. Moreover, we revealed that these spin-split defect states exhibit unidirectional spin polarization in momentum $\vec{k}$-space, which is oriented perpendicularly to the VLD orientation. Using $\vec{k}\cdot\vec{p}$ perturbation theory and symmetry analysis, we demonstrate that the 1D spin-split Rashba states with collinear spin polarization are enforced by the lowering of symmetry of the VLDs into the $C_{s}$ point group, which retains the $M_{xz}$ mirror symmetry together with the 1D nature of the VLDs. These 1D Rashba states protect carriers against spin decoherence and support an exceptionally long spin lifetime \cite{Dyakonov, Schliemann, J_Schliemann}, making them promising for developing efficient spintronic devices. Finally, we also discuss the potential applications of this system in spintronics.

\section{Computational Details}

All DFT calculations were performed based on norm-conserving pseudo-potentials and optimized pseudo-atomic localized basis functions implemented in the OpenMX code \cite{Ozaki, Ozakikino, Ozakikinoa}. The exchange-correlation function was treated within generalized gradient approximation by Perdew, Burke, and Ernzerhof (GGA-PBE) \cite{gga_pbe, Kohn}. The basis functions were expanded by a linear combination of multiple pseudo atomic orbitals (PAOs) generated using a confinement scheme \cite{Ozaki, Ozakikino}. The accuracy of the basis functions, as well as pseudo-potentials we used, were carefully bench-marked by the delta gauge method \cite{Lejaeghere}. The orbitals for Si and Bi atoms were specified by Si7.0-$s^2p^2d^1$ and Bi8.0-$s^3p^2d^2f^1$, which means that the cutoff radii of Si and Bi atoms are 7.0 and 8.0 Bohr, respectively. The cutoff energy was set to 300 Ry for the convergence of the charge density.

\begin{figure}
	\centering
		\includegraphics[width=1.0\textwidth]{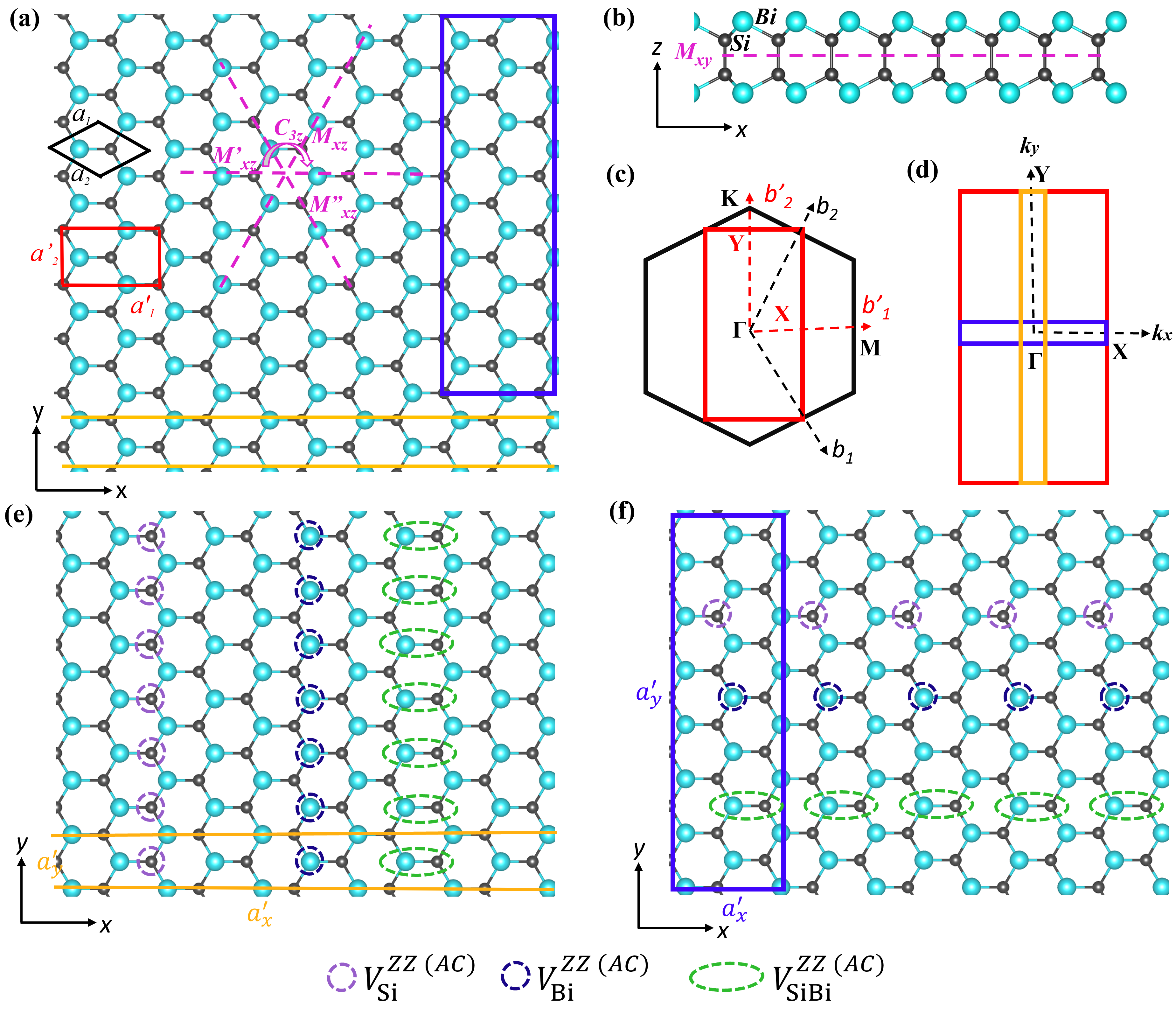}
	\caption{(a)-(b) Top and side views of the pristine Si$_2$Bi$_2$ ML are shown, respectively. The black, red, yellow, and blue lines show the primitive hexagonal, minimum rectangular cells, rectangular supercell along the armchair ($x$) direction, and rectangular supercell along the zigzag ($y$) direction, respectively. The symmetry operations of the crystal in the pristine system are indicated. (c) Folding of the first-Brillouin zone (FBZ) between primitive hexagonal and minimum rectangular cells and (d) the FBZ for the pristine supercell are given. (e) The atomic structures of the zigzag-VLD systems are depicted, with purple, black, and green dashed circles representing the $V_{\texttt{Si}}^{ZZ}$, $V_{\texttt{Bi}}^{ZZ}$, and $V_{\texttt{SiBi}}^{ZZ}$ systems, respectively. (f) The atomic structures for the armchair-VLD systems ($V_{\texttt{Si}}^{AC}$, $V_{\texttt{Bi}}^{AC}$, and $V_{\texttt{SiBi}}^{AC}$) are shown, similar to Fig. 1(e). For the zigzag (armchair)-VLD systems, line defects are oriented along the $y$ ($x$) direction, while they are isolated in the direction perpendicular to the line defect.}
	\label{figure:Figure1}
\end{figure}

Figs. 1(a)-1(b) illustrate the top and side views of the crystal structure of the pristine Si$_2$Bi$_2$ ML, respectively, exhibiting its primitive hexagonal cell, minimal rectangular cells, and rectangular supercells along both the armchair ($x$) and zigzag ($y$) directions. This corresponds to the folding of the first-Brillouin zone (FBZ) between the primitive hexagonal and minimal rectangular cells [Fig. 1(c)], as well as the FBZ for the pristine supercell [Fig. 1(d)]. The vacancy line defect (VLD) systems were modeled using a supercell approach where the minimal rectangular cell was expanded to $1\times7\times1$ along the armchair direction and $7\times1\times1$ along the zigzag direction, sufficient to prevent interactions between periodic images of line defects. We considered six different VLD configurations: (i) a single Si VLD along the zigzag direction ($V_{\texttt{Si}}^{ZZ}$), (ii) a single Bi VLD along the zigzag direction ($V_{\texttt{Bi}}^{ZZ}$), (iii) a Si-Bi VLD along the zigzag direction ($V_{\texttt{SiBi}}^{ZZ}$), (iv) a single Si VLD along the armchair direction ($V_{\texttt{Si}}^{AC}$), (v) a single Bi VLD along the armchair direction ($V_{\texttt{Bi}}^{AC}$), and (vi) a Si-Bi VLD along the armchair direction ($V_{\texttt{SiBi}}^{AC}$). The atomic structure for the zigzag-VLD systems is shown in Fig. 1(e), where the purple, black, and green dashed circles represent the $V_{\texttt{Si}}^{ZZ}$, $V_{\texttt{Bi}}^{ZZ}$, and $V_{\texttt{SiBi}}^{ZZ}$ systems, respectively, while for the armchair-VLD systems including $V_{\texttt{Si}}^{AC}$, $V_{\texttt{Bi}}^{AC}$, and $V_{\texttt{SiBi}}^{AC}$, are presented in Fig. 1(f). For the zigzag (armchair) VLD systems, the line defects are oriented along the $y$ ($x$) direction and are isolated in the perpendicular direction to the line defect. We used a periodic slab with a sufficiently large vacuum layer (20 \AA) to avoid interactions between adjacent layers. The geometries were fully relaxed until the force acting on each atom was less than 1 meV/\AA. For the zigzag and armchair-ACVLD systems, we used $6\times3\times1$ and $3\times6\times1$ $k$-point meshes, respectively.

To confirm the energetic stability of the VLD, we compute the formation energy $E^{f}$ through the following equation \cite{Freysoldt}  
\begin{equation}
\label{1}
E^{f}=E_{\texttt{VLD}}-E_{\texttt{Pristine}}+\sum_{i}n_{i}\mu_{i}.
\end{equation}
In Eq. (1), $E_{\texttt{VLD}}$ is the total energy of the VLD systems, $E_{\texttt{Pristine}}$ is the total energy of the pristine Si$_2$Bi$_2$ ML supercell, $n_{i}$ is the number of atom being removed from the pristine system, and $\mu_{i}$ is the chemical potential of the removed atoms corresponding to the chemical environment surrounding the system. The chemical potentials $\mu_{i}$ closely depend on the growth conditions and can be determined by the following relations:
\begin{equation}
\label{2}
\mu_{\texttt{Si}} + \mu_{\texttt{Bi}} = \mu_{\texttt{Si}_{2}\texttt{Bi}_{2}}      
\end{equation}
\begin{equation}
\label{3}
\mu_{\texttt{Si}}\leq \mu_{\texttt{Si}\left[\texttt{bulk}\right]}      
\end{equation}
\begin{equation}
\label{4}
\mu_{\texttt{Bi}}\leq \mu_{\texttt{Bi}\left[\texttt{bulk}\right]}      
\end{equation}
The Si$_2$Bi$_2$ ML is usually grown under Si-rich or Bi-rich conditions, which directly determine the type and concentration of native defects. Thus, we examine these two extreme scenarios to determine the defect formation energy. In the Si-rich case, $\mu_{\texttt{Si}}$ corresponds to the energy of a Si atom in its bulk phase ($\mu_{\texttt{Si}\left[\texttt{bulk}\right]}$). Likewise, for the Bi-rich condition, $\mu_{\texttt{Bi}}$ equals the energy of a Bi atom in its bulk phase ($\mu_{\texttt{Bi}\left[\texttt{bulk}\right]}$). To further confirm stability of the VLD systems, we performed ab-initio molecular dynamics (AIMD) simulations at room temperature (300 K) with a total simulation time of 4 ps.

To analyze the electronic properties of defective Si$_2$Bi$_2$ ML, we performed the band unfolding calculation within the framework of the OpenMX code \cite{CCLee_2013}. In this band unfolding method, the unfolded spectral weight is obtained by projecting the spectral function of the supercell onto the basis functions of reference (primitive) cell. The unfolded spectral weight $A_{\vec{k}j,\vec{k}j}(E)$ at specific momentum $\vec{k}$ point and band index $j$ was obtained as \cite{CCLee_2013}
\begin{equation}
\label{4a}
A_{\vec{k}j,\vec{k}j}(E)=\frac{L}{l}\sum_{\vec{K}\vec{G}} \delta_{\vec{k}-\vec{G}, \vec{K}} W_{\vec{K}J} A_{\vec{K}J,\vec{K}J}(E)      
\end{equation}
with
\begin{equation}
\label{4b}
W_{\vec{K}J}=\sum_{MNr}\exp^{i\vec{k}\cdot \left(\vec{r}-\vec{r'}(M)\right)} C_{M}^{\vec{K}J} \left(C_{M}^{\vec{K}J}\right)^{*}S_{0N,rm(M)}     
\end{equation}
where $L(l)$ is the number of the unit cells introduced in the Born-von Karman boundary condition for the supercell (primitive cell) and $A_{\vec{K}J,\vec{K}J}(E)$ is a delta function at the eigen-value $\delta(E-E_{\vec{K}J})$. In Eq. (\ref{4a}), $S_{0N,rm(M)}$ is the overlap integral evaluated by assuming the positions of LCAO basis functions is used in the actual supercell calculation. Therefore, expanding the wave function using the LCAO basis allows us to analyze the contribution of each atomic basis function to the unfolded spectral function. The orbitally decomposed spectral weights are further utilized to study the defect states in the defective Si$_2$Bi$_2$ monolayer along a specific $\vec{k}$-path in the FBZ. 

Finally, to characterize the spin-splitting properties of the defective Si$_2$Bi$_2$ ML, we calculate spin-resolved projected bands by evaluating the spin vector components ($S_{x}$, $S_{y}$, $S_{z}$) in $k$-space, using the spin density matrix, denoted as $P_{\sigma \sigma^{'}}(\vec{k},\mu)$, obtained after achieving self-consistency in the DFT calculations \cite{Kotaka}. The spin density matrix is calculated using the spinor Bloch wave function, $\Psi^{\sigma}_{\mu}(\vec{r},\vec{k})$, through the following equation:
\begin{equation}
\begin{aligned}
\label{5}
P_{\sigma \sigma^{'}}(\vec{k},\mu)=\int \Psi^{\sigma}_{\mu}(\vec{r},\vec{k})\Psi^{\sigma^{'}}_{\mu}(\vec{r},\vec{k}) d\vec{r}\\
                                  = \sum_{n}\sum_{i,j}[c^{*}_{\sigma\mu i}c_{\sigma^{'}\mu j}S_{i,j}]e^{\vec{R}_{n}\cdot\vec{k}},\\
\end{aligned}
\end{equation}
where $\Psi^{\sigma}_{\mu}(\vec{r},\vec{k})$ is obtained after self-consistency is achieved in the DFT calculation. In Eq. (\ref{1}), $S_{ij}$ is the overlap integral of the $i$-th and $j$-th localized orbitals, $c_{\sigma\mu i(j)}$ is expansion coefficient, $\sigma$ ($\sigma^{'}$) is the spin index ($\uparrow$ or $\downarrow$), $\mu$ is the band index, and $\vec{R}_{n}$ is the $n$-th lattice vector.

\section{RESULT AND DISCUSSION}

\subsection{Pristine Si$_2$Bi$_2$ ML}

Before examining the defective systems, we first provide a brief overview of the structural symmetry and electronic properties of the pristine system. The Si$_2$Bi$_2$ ML is a member of the 2D group IV-V $A_{2}X_{2}$ compounds, and its stability and electronic properties have been previously reported \cite{Ozdamar, Bafekry, Lee2020, Lukmantoro, Absor_Arif}. These compounds consist of covalently bonded quadruple atomic layers arranged in an alternating Bi-Si-Si-Bi sequence, resulting in a trigonal prismatic structure where Si atoms form a triangular prism around the Bi dimer [Figs. 1(a)-1(b)]. This structure is similar to that previously reported on GaSe ML \cite{Hirokazu}, which is reminiscent of the $1H$-phase of TMDCs MLs \cite{Zhu2011}. The Si$_2$Bi$_2$ ML possess crystal symmetry belonging to the $P\bar{6}m_{2}$ space group with a $D_{3h}$ point group, preserving the following symmetry operations: identity ($E$), in-plane mirror reflection ($M_{xy}$), out-of-plane mirror reflection ($M_{xz}$, $M_{xz}^{'}$, $M_{xz}^{"}$), two-fold rotation around the axis parallel to the $M_{xz}$ plane ($C_{2}$, $C_{2}^{'}$, $C_{2}^{"}$), three-fold rotation around the $z$-axis ($C_{3}$, $C^{2}_{3}$), and three-fold improper rotation around the $z$-axis ($S_{3}$, $S^{2}_{3}$); see Figs. 1(a)-1(b). The optimized lattice parameter in a primitive hexagonal unit cell is 4.13 \AA, aligning well with prior theoretical studies \cite{Ozdamar, Bafekry, Lee2020, Lukmantoro, Absor_Arif}.

\begin{figure*}
	\centering
		\includegraphics[width=0.9\textwidth]{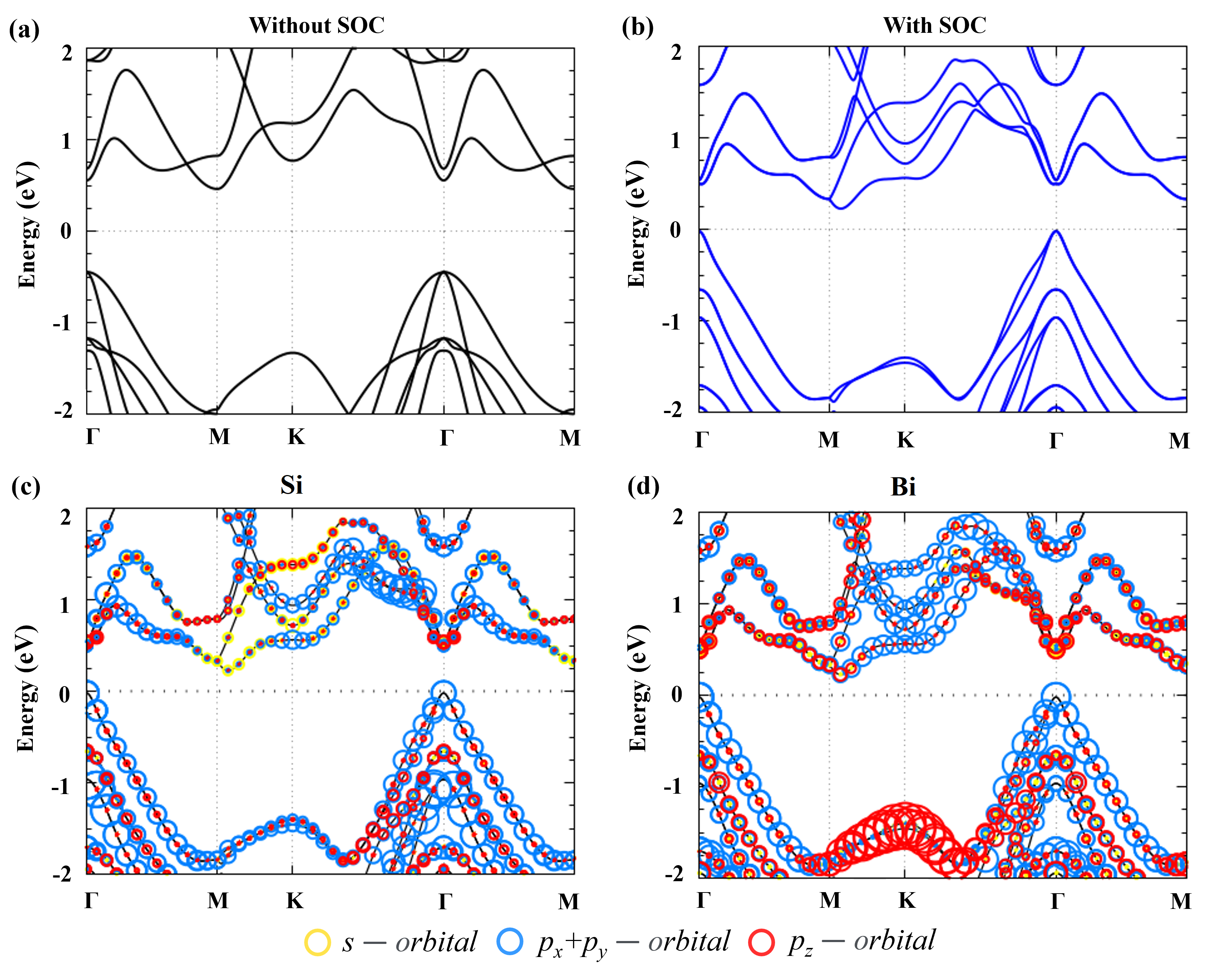}
	\caption{Band structures of the pristine Si$_{2}$Bi$_{2}$ ML in the ($1\times1$) primitive hexagonal unit cell along the selected $\vec{k}$ path in the FBZ calculated (a) with (red lines) and (b) without (black lines) the SOC are shown. Orbital-resolved electronic bands projected to the (c) Si and (d) Bi atoms. The orbital atomic components are indicated by the color circles, where the radii of the circles reflect the magnitude of the spectral weight of the particular orbitals to the bands.}
	\label{figure:Figure2}
\end{figure*}

Figs. 2(a)-2(b) present the electronic band structure of the pristine Si$_{2}$Bi$_{2}$ ML within the ($1\times1$) primitive hexagonal unit cell along a specific $\vec{k}$-path in the FBZ, calculated both with (blue lines) and without (black lines) SOC, respectively. This corresponds to the orbital-resolved electronic bands projected onto the Si and Bi atoms shown in Figs. 2(c)-(d), respectively. Without the SOC, it is found that the Si$_{2}$Bi$_{2}$ ML is an indirect semiconductor with its valence band maximum (VBM) at the $\Gamma$ point and conduction band minimum (CBM) at the $M$ point [Fig. 2(a)]. The calculated band gap is 1.2 eV under GGA level, which is in a good agreement with previous studies \cite{Ozdamar, Bafekry, Lee2020, Lukmantoro, Absor_Arif}. The indirect band gap is preserved, but the band gap decreases to 0.28 eV when the SOC is included in the calculation [Fig. 2(b)]. This reduction is attributed to the upward shift of the VBM at the $K$ point, making it higher in energy, and the downward shift of the CBM to a location between the $M$ and $K$ points [Fig. 2(b)]. These energy shifts of the VBM and CBM occur, which is due to the strong orbital hybridization between the Si and Bi atoms. In fact, our analysis of the band projections onto the atomic orbitals confirms that the CBM of the Si$_{2}$Bi$_{2}$ ML ML predominantly arises from strong admixtures between Bi-$p_{x}+p_{y}$ and Bi-$p_{z}$ orbitals with small contribution of the Si-$s$ and Si-$p_{z}$ orbitals, while the VBM is mainly originated from the contribution of the Bi-$p_{x}+p_{y}$ and Si-$p_{x}+p_{y}$ orbitals [Figs. 2(c)-(d)]. Additionally, SOC induces significant band splitting in both the CBM and VBM due to the inversion symmetry breaking of the crystal, especially near the band edges close to the Fermi level, except at time-reversal-invariant $k$ points like $\Gamma$ and $M$, and the high symmetry line $\Gamma-M$ [Fig. 2(b)]. Consistent with previous studies \cite{Lukmantoro, Absor_Arif}, the spin splitting of up to 0.72 eV is observed at near the CBM along the $M-K$ line, which is larger than that observed in several 2D transition metal dichalcogenide (TMDC) monolayers [0.15 eV – 0.55 eV] \cite{Zhu2011, Absor2016, Absor2017}.

\subsection{Vacancy line defects in Si$_2$Bi$_2$ ML}
\subsubsection{Structural stability and symmetry}

When a VLD is introduced into the Si$_{2}$Bi$_{2}$ ML, notable changes in atomic structures occur due to atomic relaxation. To evaluate the energetic stability of the VLD systems, Table I presents the calculated formation energies of the defective systems under Si-rich and Bi-rich conditions. We find that the $V_{\texttt{Si}}^{ZZ}$ has the lowest formation energies in both conditions, indicating that the $V_{\texttt{Si}}^{ZZ}$ is the most stable VLD systems formed in the Si$_{2}$Bi$_{2}$ ML. This stability aligns with previous findings that group IV vacancy defects form easily in 2D group IV-V compounds, as previously reported for In$_{2}$Se$_{2}$ \cite{Xiao2017, WangRSC, Wang2018} and  Si$_{2}$P$_{2}$ \cite{Ha2024, Chen2015} MLs. Conversely, the formation of other defective systems ($V_{\texttt{Bi}}^{ZZ}$, $V_{\texttt{SiBi}}^{ZZ}$, $V_{\texttt{Si}}^{AC}$, $V_{\texttt{Bi}}^{AC}$, $V_{\texttt{SiBi}}^{AC}$) is less favorable due to the required electron energy. Since the Bi atom is covalently bonded to three neighboring Si atoms, removing Bi atoms disrupts the Si sublattice, and hence raising the formation energy; see Table I. We emphasized here that our calculations did not account for lattice shrinkage and stretching, which can effectively alter the formation energy and engineer different types of VLDs \cite{Fang2019, Chen2022}. Moreover, we also suggest that forming $V_{\texttt{Si}}^{ZZ}$ in the Si$_{2}$Bi$_{2}$ ML is more favorable under Si-rich conditions than Bi-rich conditions due to the lower formation energy in Si-rich conditions, as shown in Table I.

\begin{table}[ht!]
\caption{The calculated formation energy (measured in eV) of the VLD systems under the Si-rich ($E^{f}_{\texttt{Si-rich}}$) and Bi-rich ($E^{f}_{\texttt{Bi-rich}}$) conditions is presented. The calculated Si-Bi and Si-Si bond lengths (measured in \AA) for the atoms located parallel ($d^{\|}_{\texttt{Si-Bi}}$, $d^{\|}_{\texttt{Si-Si}}$) and perpendicular ($d^{\bot}_{\texttt{Si-Bi}}$, $d^{\bot}_{\texttt{Si-Si}}$) to the SiBi plane near the VLD site, are shown. } 
\centering 
\begin{tabular}{cc cc cc cc cc cc cc cc} 
\hline\hline 
 Defective systems && ($E^{f}_{\texttt{Si-rich}}$; $E^{f}_{\texttt{Bi-rich}}$) && $d^{\|}_{\textbf{Si-Bi}}$ && $d^{\|}_{\textbf{Si-Si}}$ && $d^{\bot}_{\textbf{Si-Bi}}$   &&  $d^{\bot}_{\textbf{Si-Si}}$     \\ %
\hline %
\textbf{Pristine}                            &&                && 2.74        &&        &&           &&   2.36           \\
\textbf{Zigzag VLD}                          &&                &&             &&        &&           &&                   \\
 $V_{\texttt{Si}}^{ZZ}$                      && (0.34; 0.99)   && 2.75 - 3.01 &&        && 2.72      &&  2.34 - 2.38     \\
 $V_{\texttt{Bi}}^{ZZ}$                      && (3.47; 2.82)   && 2.74 - 2.86 &&        &&           &&  2.34 - 2.37      \\ 
 $V_{\texttt{SiBi}}^{ZZ}$                    && (1.90; 1.90)   && 2.71 - 2.81 &&        && 2.76      &&  2.33 - 2.38      \\
\textbf{Armchair VLD}        								 &&                &&             &&        &&           &&                 \\
 $V_{\texttt{Si}}^{AC}$                      && (0.61; 1.27)   && 2.75 - 2.76 &&        &&           &&  2.35              \\
 $V_{\texttt{Bi}}^{AC}$                      && (2.69; 2.00)   && 2.74        && 2.57   &&           &&  2.37                \\ 
 $V_{\texttt{SiBi}}^{AC}$                    && (2.45; 2.45)   && 2.90        && 2.55   &&           &&  2.35 - 2.39         \\       
\hline\hline 
\end{tabular}
\label{table:Table 1} 
\end{table}

\begin{figure*} 
	\centering
		\includegraphics[width=1.0\textwidth]{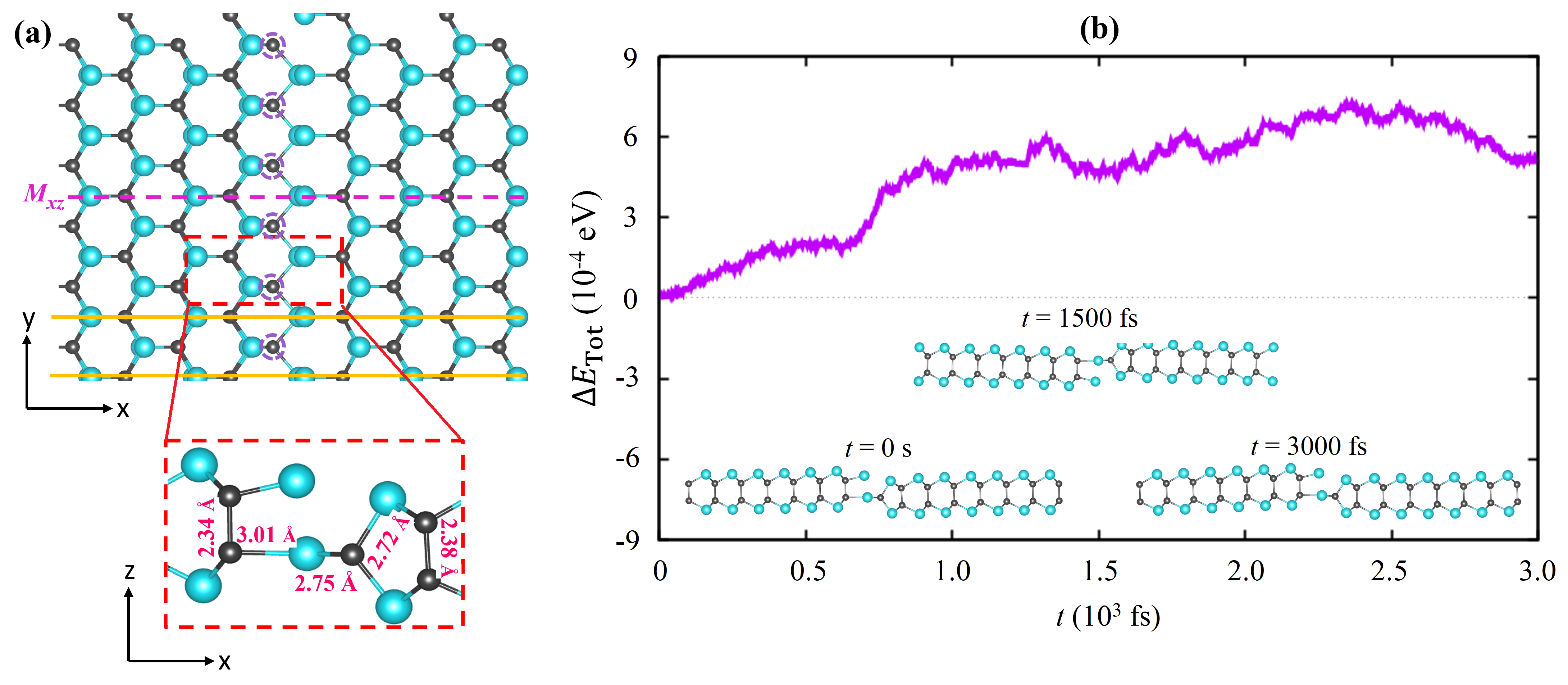}
	\caption{(a) The optimized geometry of the $V_{\texttt{Si}}^{ZZ}$ is shown, where region around the VLD sites are highlighted with red dashed lines. The bondlengths between the atoms near the VLD site are indicated. (b) The result of the AIMD simulation of the $V_{\texttt{Si}}^{ZZ}$ is shown by the different total energy ($\Delta E_{\texttt{Tot}}$) relative to the initial structures, plotted as a function of simulation time $t$. Snapshots of the geometry at $t=0$ s, $t=1500$ fs, and $t=3000$ fs are highlighted. }
	\label{figure:Figure3}
\end{figure*}

Relaxation causes slight shifts in the atomic positions around the VLD site from their original locations, as reflected in the changes in atomic bond lengths. These calculated bond lengths for all VLD systems are detailed in Table I. Focusing on the $V_{\texttt{Si}}^{ZZ}$ as the stablest VLD systems in the Si$_{2}$Bi$_{2}$ ML, we identify structural deformation occurred both in the SiBi plane and perpendicularly [Fig. 3(a)]. Here, three Bi atoms near the VLD site move closer together in the SiBi plane, resulting in Si-Bi bond lengths ($d^{\|}_{\textbf{Si-Bi}}$) varying from 2.75 to 3.01 \AA, which is considerably larger than the pristine ML (2.74 \AA). The Si atom beneath the $V_{\texttt{Si}}^{ZZ}$ shifts down towards the fourth sub-layer, with vertical Si-Si bond lengths ($d^{\bot}_{\textbf{Si-Si}}$) ranging from 2.34 \AA\ to 2.38 \AA\ and Si-Bi bond lengths ($d^{\bot}_{\textbf{Si-Bi}}$) of 2.72 \AA. Consequently, only the out-of-plane $M_{xz}$ mirror symmetry perpendicular to the defect line remains, as shown by the magenta dashed lines in Fig. 3(a), resulting in that the symmetry of $V_{\texttt{Si}}^{ZZ}$ belongs to the $C_{s}$ point group. To further investigate how its structural symmetry and geometry evolve during relaxation, we performed AIMD simulations at 300 K for the $V_{\texttt{Si}}^{ZZ}$ system. As shown in Fig. 3(b), we find that after heating for 4 ps at 300 K, the structure of the $V_{\texttt{Si}}^{ZZ}$ remains intact without any signs of disorder, indicating that the $V_{\texttt{Si}}^{ZZ}$ is thermally stable at room temperature. 

We note that, compared to the $V_{\texttt{Si}}^{ZZ}$ case, the optimized structures of other VLD systems exhibit distinct atomic relaxations and distortions, as shown in Fig. S1 of the Supplementary Materials \cite{Supplementary}. These variations are reflected in the atomic bond lengths, detailed in Table I. Nonetheless, similar to the $V_{\texttt{Si}}^{ZZ}$ case, all other VLD systems maintain a $C_{s}$ point group symmetry due to the presence of a single mirror symmetry plane ($M_{xz}$) parallel to the extended vacancy line (see Fig. S1 in the Supplementary Materials \cite{Supplementary}). This symmetry reduction to the $C_{s}$ point group, observed across all VLD systems in the Si$_{2}$Bi$_{2}$ ML, aligns with recent reports of VLD systems in various 2D TMDCS MLs \cite{Li2019, Adhib, Absor2024}.

\subsubsection{Electronic and spin-splitting properties}

Significant changes in the electronic properties of the Si$_{2}$Bi$_{2}$ ML are expected when introducing the VLD. Fig. 4(a) shows the unfolded band structures of the $V_{\texttt{Si}}^{ZZ}$, calculated with including the SOC, represented by intensity map of the sum of all the unfolded total spectral weight. It is clearly shown that the overall pristine band structure shown in Fig. S2 in the Supplementary Materials \cite{Supplementary}, is well maintained in the unfolding bands, accompanied with additional interesting features of the bands. The notable feature is the newly revealed metallic defect state bands, labeled as DS-1 and DS-2, which appear within the fundamental band gap and intersect the Fermi level [Fig. 4(a)]. These metallic defect states are positioned closer to the CBM than the VBM, indicating a higher likelihood of electron acceptance. Importantly, we identify large spin-splitting bands in the DS-1 and DS-2 defect states except at the time-reversal invariant points ($\Gamma$, $Y$), having the maximum spin-splitting energy close to the Fermi level highlighted by $\Delta E_{1}$ and $\Delta E_{2}$, respectively; see Fig. 4(b). Our orbital resolved bands projected to the Si and Bi atoms near the VLD sites confirmed that these spin-split defect  states are mainly driven by admixtures of the $s$ and $p_{x+y}$ orbitals of the Si atoms and $s$, $p_{x+y}$, and $p_{z}$ orbitals of the Bi atoms near the VLD sites [Figs. 4(c)-4(d)]. 

\begin{figure*} 
	\centering
		\includegraphics[width=1.0\textwidth]{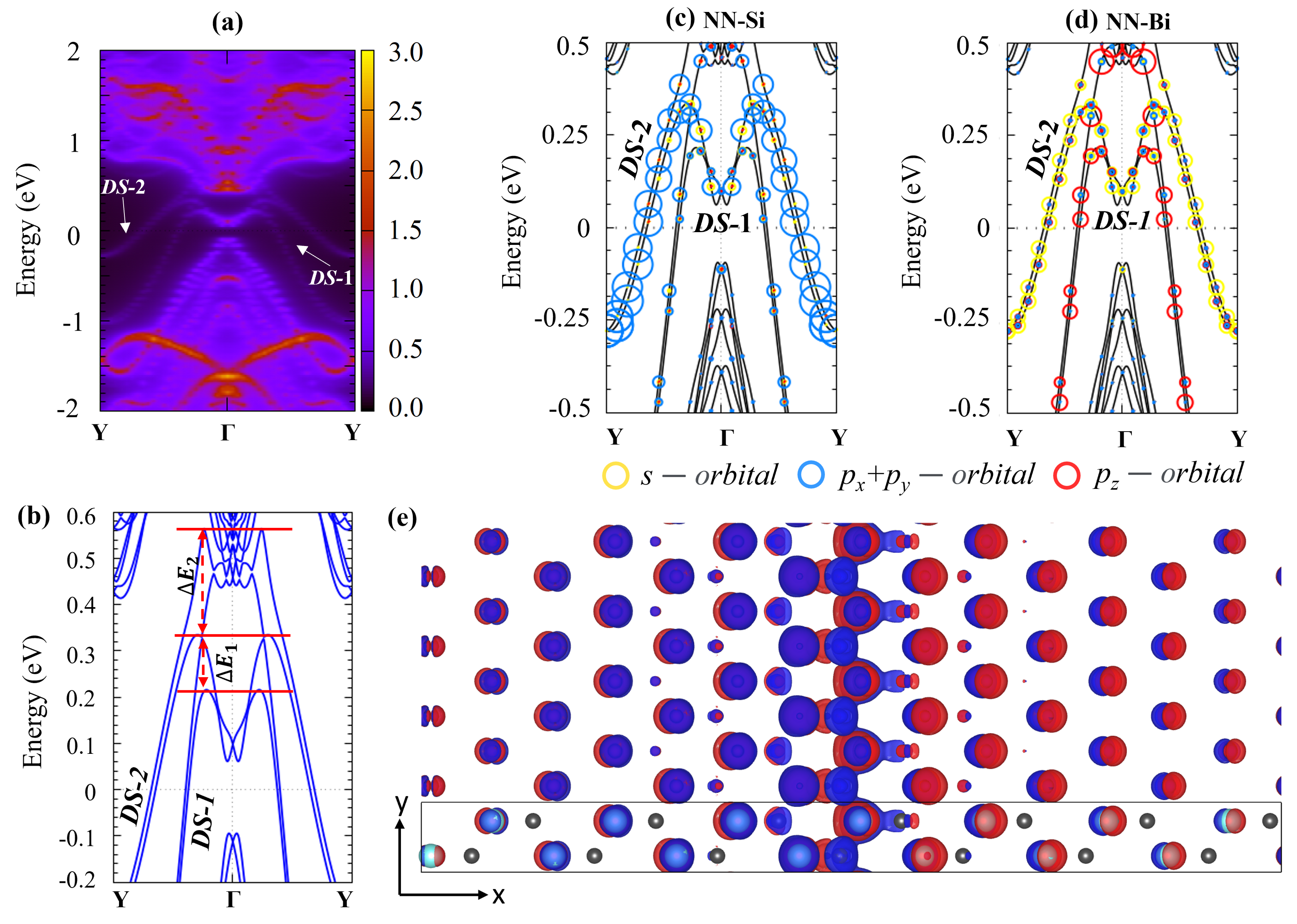}
	\caption{(a) Unfolding band of the $V_{\texttt{Si}}^{ZZ}$ calculated with including the spin-orbit coupling (SOC) presented by the sum of all the unfolded total spectral weight shown by intensity map. The spin-split defect states bands labeled by DS-1 and DS-2 are indicated. (b) The spin-split bands of the defect states (DS-1, DS-2), highlighting the maximum spin-splitting energies ($\Delta E_{1}$, $\Delta E_{2}$) near the Fermi level, are shown.(c-d) Orbital-resolved projected bands projected to the atoms near the VLD sites at the nearest neighbor (NN) Si and Bi atoms, respectively, are presented. Two spin-split metallic defect states near the Fermi level are labeled by DS-1 and DS-2. The orbital-resolved projected bands are presented by the color circle, where the radii of the circles reflect the magnitude of the spectral weight of the particular orbitals to the bands. (e) The isosurface map of the charge density difference calculated for $V_{\texttt{Si}}^{ZZ}$ is shown, with an isovalue of 0.028 used for the electron density difference.} 
	\label{figure:Figure4}
\end{figure*}

The appearance of spin-split defect states is also observed in other VLD systems, as shown by the calculated unfolded bands in Fig. S3 of the Supplementary Materials \cite{Supplementary}. With the exception of the $V_{\texttt{Si}}^{AC}$ system, which exhibits spin-split insulating defect states, other VLD systems [$V_{\texttt{Bi}}^{ZZ}$, $V_{\texttt{SiBi}}^{ZZ}$, $V_{\texttt{Bi}}^{AC}$, $V_{\texttt{SiBi}}^{AC}$] demonstrate spin-split metallic defect states. In zigzag-VLD systems [$V_{\texttt{Bi}}^{ZZ}$, $V_{\texttt{SiBi}}^{ZZ}$], these spin-split defect states predominantly arise from the hybridization of Si-$p_{x+y}$, Si-$p_{z}$, and Bi-$p_{x+y}$ orbitals, as illustrated in Fig. S4 of the Supplementary Materials \cite{Supplementary}. For armchair VLD systems [$V_{\texttt{Si}}^{AC}$, $V_{\texttt{Bi}}^{AC}$, $V_{\texttt{SiBi}}^{AC}$], the defect states are primarily contributed by the hybridization of Si-$p_{x+y}$, Si-$p_{z}$, Bi-$p_{x+y}$, and Bi-$p_{z}$ orbitals, as detailed in Fig. S5 of the Supplementary Materials \cite{Supplementary}. Remarkably, the strong $p$-$p$ orbital coupling between Si and Bi atoms near the VLD site plays a key role in inducing significant spin splitting in the defect states, similar to phenomena reported in various 1D systems \cite{Mihalyuk, Qin, Naumov}.

The role of the $p-p$ coupling orbitals inducing the large spin splitting in the defect states can be understood by considering the nonzero SOC matrix element driven by the coupling of the atomic orbitals. Here, the SOC matrix element is proportional to $\xi_{l}\left\langle \vec{L}\cdot\vec{S}\right\rangle_{u,v}$, where $\xi_{l}$ is angular momentum resolved atomic SOC strength with $l=(s,p,d)$, $\vec{L}$ and $\vec{S}$ are the orbital angular momentum and Pauli spin operators, respectively, and $(u,v)$ is the atomic orbitals. Therefore,  only the orbitals with a nonzero magnetic quantum number ($m_{l}\neq 0$) will contribute significantly to the spin splitting. This aligns with our observation that the in-plane $p-p$ coupling orbitals ($m_{l}=\pm 1$) cause significant spin splitting, as demonstrated by the spin-split defect states (DS-1, DS-2) shown in in Figs. 4(c)-4(d). 

We highlight that the spin-split defect states in VLD systems are localized along the $\vec{k}$-direction, which is parallel to the extended defect line. In the $V_{\texttt{Si}}^{ZZ}$ system, where the defect line runs along the zigzag ($y$) direction, these defect states are notably confined along the $\Gamma-Y$ direction [Fig. 4(e)]. This confinement arises from the strong interactions between neighboring Si and Bi atoms near the VLD site along the extended vacancy line ($y$-direction), as further supported by the orbital hybridization depicted in Figs. 4(c)-(d). Our calculated charge densities confirm this localization, revealing a quasi-1D conducting channel that restricts electron motion to the extended defect line [Fig. 4(e)]. In contrast, for armchair-VLD systems, localized defect states are observed along the $\Gamma-X$ direction, as demonstrated by the $V_{\texttt{Si}}^{AC}$ system depicted in Figs. S6(a)-(c) of the Supplementary Materials \cite{Supplementary}. Notably, this 1D confinement of defect states is expected to enable strong band-like charge transport with significantly improved carrier mobilities \cite{Fishchuk, Gupta}, that is beneficial for transport-based spintronic devices.

\begin{table}[ht!]
\caption{The maximum spin-splitting energy for spin-split defect states, denoted as $\Delta E_{1}$ (eV) and $\Delta E_{2}$ (eV), has been calculated for all VLD systems and compared with the values observed in some selected 1D Rashba materials.} 
\centering 
\begin{tabular}{cccc cccc cccc cccc} 
\hline\hline 
  Systems &&&&  $\Delta E_{1}$ (eV)  &&&&  $\Delta E_{2}$ (eV) &&&&  Reference \\ 
\hline 
 \textbf{VLD in Si$_{2}$Bi$_{2}$ ML}        &&&&                &&&&              &&&&                                 \\
 $V_{\texttt{Si}}^{ZZ}$                     &&&&     0.12       &&&&  0.22        &&&&         This work    \\
 $V_{\texttt{Bi}}^{ZZ}$                     &&&&     0.13       &&&&  0.15        &&&&        This work    \\
 $V_{\texttt{SiBi}}^{ZZ}$                   &&&&     0.17       &&&&  0.10         &&&&           This work    \\
 $V_{\texttt{Si}}^{AC}$                     &&&&     0.12       &&&&  0.04        &&&&           This work    \\
 $V_{\texttt{Bi}}^{AC}$                     &&&&     0.10       &&&&  0.17         &&&&      This work    \\
 $V_{\texttt{SiBi}}^{AC}$                   &&&&     0.10       &&&&  0.12         &&&&            This work    \\
 VLD on $1T'$ WTe$_{2}$ ML                  &&&&  0.037 - 0.14  &&&&  0.04 - 0.07 &&&&            Ref. \cite{Absor2024}    \\
 1D Pb/Si(100)                              &&&&  0.018 - 0.069 &&&&              &&&&              Ref. \cite{Mihalyuk}    \\
 1D Pt/Si(110)                              &&&&  0.081         &&&&              &&&&             Ref. \cite{Park}    \\
 Bi-adsorbed 1D In chains                   &&&&  0.059 - 0.139 &&&&              &&&&                Ref. \cite{Tanaka}    \\
\hline\hline 
\end{tabular}
\label{table:Table 2} 
\end{table}

\begin{figure*}
	\centering
		\includegraphics[width=1.0\textwidth]{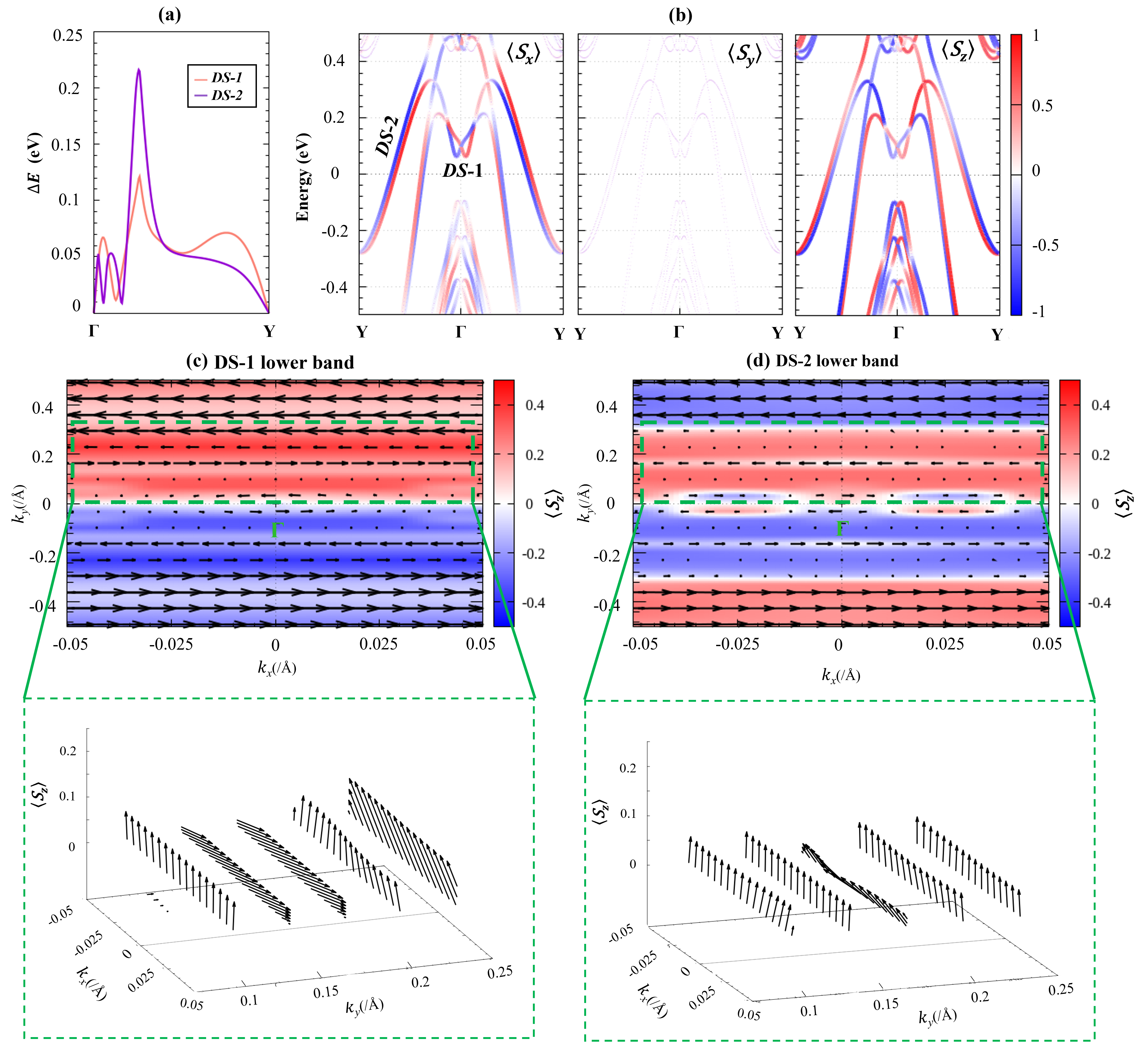}
	\caption{(a) The calculated spin-splitting energy of the $V_{\texttt{Si}}^{ZZ}$ systems for the DS-1 and DS-2 evaluated by $\Delta E=\left|E(k,\uparrow)-E(k,\downarrow)\right|$, where $E(k,\uparrow)$ ($E(k,\downarrow)$) is the energy bands at the certain $\vec{k}$ point with up (down) spin polarizations. (b) Spin-resolved projected bands calculated for DS-1 and DS-2 with $S_{x}$, $S_{y}$, and $S_{z}$ spin components are shown. $\left\langle S_{x}\right\rangle$, $\left\langle S_{y}\right\rangle$, and $\left\langle S_{z}\right\rangle$ represent the expected values of the spin components as indicated by the color bar. Spin textures projected to the FBZ for the lower bands of DS-1 (c) and DS-2 (d) are shown. The black arrows indicate the orientation of the in-plane ($S_{x}$, $S_{y}$) spin components, while the color scales represent the $S_{z}$ spin components. The insets in Figs. 5(c)-(d) display the 3D plots of the spin textures for the DS-1 and DS-2 bands, respectively, at specific $\vec{k}$ surfaces marked by the green dashed lines.}
	\label{figure:Figure5}
\end{figure*}

To analyze the spin-split defect states, we calculate the spin-splitting energy, $\Delta E(k)$, which is determined using the relation $\Delta E(k) = \left| E(k,\uparrow) - E(k,\downarrow) \right|$, where $E(k,\uparrow)$ and $E(k,\downarrow)$ represent the energy bands at a given $\vec{k}$ point with spin-up and spin-down polarizations, respectively. Figure 5(a) shows the $k$-dependent spin-splitting energy $\Delta E(k)$ of the $V_{\texttt{Si}}^{ZZ}$ system, calculated at the defect states (DS-1, DS-2) along the $\Gamma-Y$ direction. The analysis reveals that both DS-1 and DS-2 reach their maximum spin-splitting energy ($\Delta E_{1}$, $\Delta E_{2}$) at the same $k$ point, located near the $\Gamma$ point. The maximum spin-splitting energies ($\Delta E_{1}$, $\Delta E_{2}$) for all VLD systems are summarized in Table II. All the VLD systems exhibit comparable maximum spin-splitting energies ($\Delta E_{1}$ and $\Delta E_{2}$), with the exception of $V_{\texttt{Si}}^{AC}$, which exhibits the smallest $\Delta E_{2}$. Specifically, for the $V_{\texttt{Si}}^{ZZ}$ system, the maximum spin-splitting energies are $\Delta E_{1} = 0.12$ eV and $\Delta E_{2} = 0.22$ eV for DS-1 and DS-2, respectively. These values are significantly larger than those reported for other VLD systems, such as the $1T'$ phase of WTe$_{2}$ monolayers (0.037–0.14 eV) \cite{Absor2024}, 1D Pb/Si(100) (0.018–0.069 eV) \cite{Mihalyuk}, 1D Pt/Si(110) (0.081 eV) \cite{Park}, and Bi-adsorbed 1D In chains (0.059–0.139 eV) \cite{Tanaka}.

To further explore the nature of the spin-split defect states, Fig. 5(b) displays the spin-resolved projected bands of the $V_{\texttt{Si}}^{ZZ}$ system along the $\Gamma-Y$ line, emphasizing the expectation values of the $S_{x}$, $S_{y}$, and $S_{z}$ spin components. The analysis shows that the defect states DS-1 and DS-2 exhibit spin polarizations primarily dominated by the $S_{x}$ and $S_{z}$ components along the $\Gamma-Y$ path, while the $S_{y}$ component is nearly zero. This indicates that the spin polarizations are tilted within the $x-z$ plane in the FBZ. Symmetrically, the presence of the $M_{xz}$ mirror plane in the $V_{\texttt{Si}}^{ZZ}$ system implies that the $S_{y}$ spin component should ideally be zero, which aligns with the spin-split defect states (DS-1 and DS-2) illustrated in Fig. 5(b). However, in our computational results, we observed nonzero $S_{y}$ components, as represented by the faint-colored lines indicating the $\left\langle S_{y}\right\rangle$ values in Fig. 5(b). These values, nevertheless, are extremely small, on the order of $10^{-6}$, when compared to the $\left\langle S_{x}\right\rangle$ and $\left\langle S_{z}\right\rangle$ values. This minor discrepancy is likely due to the inherent limitations in computational accuracy often encountered in such calculations.

Furthermore, by projecting the spin polarization onto the FBZ plane, it is observed that the spin polarizations are aligned unidirectionally along the $k_{x}$ direction and are tilted out of the plane along the $k_{z}$ direction; see Figs. 5(c) and 5(d). This collinear tilted spin polarization along the $x-z$ plane, enforced by the out-of-plane mirror symmetry $M_{xz}$, results in the so-called canted persistent spin textures (PST). This unique PST significantly differs from previously reported PSTs in VLD-based 2D TMDC systems \cite{Li2019, Adhib, Absor2024}. Notably, similar canted PSTs have been reported on the ZnO (10$\overline{1}$0) surface \cite{Absor_2015} and in 2D $1T'$ phase TMDCs, such as W(Mo)Te$_{2}$ MLs \cite{Garcia2020, Vila2021, AbsorJAP2022}. The observed canted PST in the current VLD system is anticipated to generate a unidirectional canted spin-orbit field in $k$-space, which protects spin coherence and leads to an extremely long spin lifetime \cite{Dyakonov, Schliemann, J_Schliemann}. In fact, robust spin diffusion over long distances driven by canted PST has been experimentally demonstrated in few-layer MoTe$_{2}$ at room temperature \cite{Song2020}, making this system a promising candidate for efficient spintronic applications.

We emphasize that, despite both the zigzag- and armchair-VLD systems sharing the same point group symmetry ($C_{s}$), they exhibit distinctly different spin polarization behaviors. This difference stems from the unique characteristics of the 1D localized spin-split defect states in each VLD system. For example, in the $V_{\texttt{Si}}^{AC}$ system, representing the armchair-VLD systems, localized spin-split defect states (DS-1, DS-2) with significant spin-splitting energy are observed along the $\Gamma-X$ ($k_{x}$) direction; see Figs. S6(a)-(b) in the Supplementary Materials \cite{Supplementary}. This behavior contrasts with that of the $V_{\texttt{Si}}^{ZZ}$ system. The localized charge density along the extended defect line, as shown in Fig. S6(c) of the Supplementary Materials \cite{Supplementary}, further confirms the 1D nature of these spin-split defect states. Consequently, only the $S_{y}$ spin component is maintained in the spin-split defect states, which results in a unidirectional spin polarization along the $k_{y}$ direction in the FBZ; see Figs. S6(d)-(f) in the Supplementary Materials \cite{Supplementary}.

\subsubsection{Symmetry-based $\vec{k}\cdot\vec{p}$ Hamiltonian model}

To understand the observed spin-splitting and spin polarization in the spin-split defect states of the VLD systems, we developed a $\vec{k}\cdot\vec{p}$ Hamiltonian model based on group theory analysis. Due to the preservation of time-reversal symmetry in the VLD systems, Kramer's degeneracy is maintained at the high-symmetry $\vec{k}$-points in the FBZ, such as the $\Gamma$, $X$, and $Y$ points. However, at $\vec{k}$-points away from these time-reversal-invariant points, Kramer's doublet is disrupted due to the presence of SOC, which can be described using a $\vec{k}\cdot\vec{p}$ Hamiltonian. To further derive the effective SOC Hamiltonian $\hat{H}_{\textbf{SOC}}$, we first focus on the vicinity of the $\Gamma$ point and then extend our analysis to all paths along the $\Gamma-X$ or $\Gamma-Y$ lines. Following the method developed by Vajna. $et$. $al$., $\hat{H}_{\textbf{SOC}}$ can be derived through the following invariance formulation \citep{Vajna}: 
\begin{equation}
\label{6}
H_{\textbf{SOC}}(\vec{k})= \alpha\left(g\vec{k}\right)\cdot \left(\det(g)g\vec{\sigma} \right),
\end{equation}
where $\vec{k}$ and $\vec{\sigma}$ are the electron's wave vector and spin vector, respectively, and $\alpha\left(g\vec{k}\right)=\det(g)g\alpha\left(\vec{k}\right)$, where $g$ is the element of the point group characterizing the small group wave vector $G_{\vec{K}}$ of the high symmetry point $\vec{K}$ in the first Brillouin zone. By classifying the components of $\vec{k}$ and $\vec{\sigma}$ according to the irreducible representation (IR) of $G_{\vec{K}}$, we can further decompose their direct product into IRs. Based on Eq. (\ref{6}), only the fully symmetric IR from this decomposition contributes to $\hat{H}_{\textbf{SOC}}$. Thus, utilizing the corresponding point group tables, one can readily construct the possible terms of $\hat{H}_{\textbf{SOC}}$.

\begin{table}[ht!]
\caption{Character tabel for $C_{s}$ point group with two elements of the symmetry operation, namely identity operation $E:(x,y,z)\rightarrow(x,y,z)$ and mirror symmetry operation $M_{xz}: (x,y,z)\rightarrow(x,-y,z))$. Two one-dimensional irreducible representations (IR) are shown.} 
\centering 
\begin{tabular}{c c c c c c c c c c c c c c c c c c c c c c c c c}
\hline\hline 
IRs &  & & & &  &  &  & $E$ &  &  &  &  &  &  &  & $M_{yz}$   &   & &  &  &  &   &    &   Linear, Rotation \\ 
\hline 
$A'$ &  &  &  &  &  &  &  & 1 &  &  &  &   &  &  &  & 1   &  &  &  &  &   &   &    &   $x$, $z$, $R_{y}$\\
$A"$ &  & & & &  &  &  & 1 &  & & & &  &  &  & -1   & & & &  &   &   &    &  $y$, $R_{x}$,$ R_{z}$\\
\hline\hline 
\end{tabular}
\label{table:Table III} 
\end{table}

\begin{table}[ht!]
\caption{Direct product table of the irreducible representations for $C_{s}$ point group.} 
\centering 
\begin{tabular}{c c c c c c c c c c c c c c c c c c c } 
\hline\hline 
 &&&& & & & & &$A'$ &&&& & & & & & $A"$   \\ 
\hline 
$A'$ &&&&& & & & & $A'$ & & & & &&&&& $A"$   \\
$A"$ &&&& & & & & &$A"$ & & & & &&&&& $A'$  \\ 
\hline\hline 
\end{tabular}
\label{table:Table IV} 
\end{table}

As mentioned earlier, both the zigzag- and armchair-VLD systems belong to the $C_{s}$ point group with only two symmetry elements: identity operation $E:(x,y,z)\rightarrow(x,y,z)$ and mirror symmetry operation $M_{xz}: (x,y,z)\rightarrow(x,-y,z))$. Accordingly, there are two one-dimensional IRs, namely $A'$ and $A"$ as indicated by character table shown in Tables III. Since $\vec{k}$ and $\vec{\sigma}$ can be transformed as polar and axial vectors, respectively, a comparison with the character table allow us to sort out the components of these vectors according to the IR as $A'$: $k_{x}$, $k_{z}$, $\sigma_{y}$ and $A"$: $k_{y}$, $\sigma_{x}$, $\sigma_{z}$. Moreover, from the corresponding table of direct products given in Table IV, we obtain the third order terms of $\vec{k}$ as $A'$: $k^{3}_{x}$, $k^{3}_{z}$, $k^{2}_{x}k_{z}$, $k^{2}_{y}k_{z}$, $k^{2}_{z}k_{x}$  and $A"$: $k^{3}_{y}$, $k^{2}_{x}k_{y}$, $k_{x}k_{y}k_{z}$, $k^{2}_{z}k_{y}$. By applying the table of direct products, the possible combination of the $\vec{k}$ and $\vec{\sigma}$ components up to $n$th-order in $\vec{k}$ can be obtained. Collecting all these possible terms, the effective SOC Hamiltonian $\hat{H}_{\textbf{SOC}}$ up to $n$th-order in $\vec{k}$ near the $\Gamma$ point can be written as,  
\begin{equation}
\begin{aligned}
\label{7}
H_{\textbf{SOC}}(\vec{k})= \alpha_{1}k_{x}\sigma_{y}+ \beta_{1}k_{y}\sigma_{x}+\gamma_{1}k_{y}\sigma_{z}+\delta_{1}k_{z}\sigma_{y}\\
+\alpha_{3}k^{3}_{x}\sigma_{y} +\beta_{3}k^{3}_{y}\sigma_{x}+\gamma_{3}k^{3}_{y}\sigma_{z}+ \delta_{3}k^{3}_{z}\sigma_{z}\\
+...+\alpha_{n}k^{n}_{x}\sigma_{y} + \beta_{n}k^{n}_{y}\sigma_{x} +\gamma_{n}k^{n}_{y}\sigma_{z}+ \delta_{n}k^{n}_{z}\sigma_{y},
\end{aligned}
\end{equation}
where $\alpha_{n}$, $\beta_{n}$, $\gamma_{n}$, and $\delta_{n}$ are the $n$th-order in $\vec{k}$ of the SOC strength parameters. 

For the case of the zigzag-VLD systems, the present of the 1D confined defect states along the extended defect line ($\Gamma-Y$ ($k_{y}$) direction), see Fig. 6(c), enforced all the terms containing $k_{x}$ and $k_{z}$ components vanishes. Therefore, Eq. (\ref{7}) can be rewritten as
\begin{equation}
\label{8}
H_{\textbf{SOC}}^{ZZ}(\vec{k})= \beta_{1}k_{y}\sigma_{x}+\gamma_{1}k_{y}\sigma_{z} + \beta_{3}k^{3}_{y}\sigma_{x} + \gamma_{3}k^{3}_{y}\sigma_{z} +..+\beta_{n}k^{n}_{y}\sigma_{x}+\gamma_{n}k^{n}_{y}\sigma_{z}
\end{equation}
Solving the eigenvalue problem involving the Hamiltonian of Eq. (\ref{8}) up to the third order term in $\vec{k}$, we find that
\begin{equation}
\label{9}
E_{\pm}^{ZZ}(\vec{k})=E_{0}(\vec{k}) \pm \lambda_{1}k_{y} \pm \lambda_{3} k^{3}_{y},
\end{equation}
where $E_{0}(\vec{k})=(\hbar^{2}k^{2}_{y})/2m^{*}_{y}$ is the nearly free electron energy, $m^{*}_{y}$ is the effective mass along the $\Gamma-Y$ ($k_{y}$) line, $\lambda_{1}=\sqrt{\beta^{2}_{1}+\gamma^{2}_{1}}$ and $\lambda_{3}=\sqrt{\beta^{2}_{3}+\gamma^{2}_{3}}$  are the first and third order SOC strength parameters, respectively, evaluated for the spin-split defect states along the $\Gamma-Y$ ($k_{y}$) line. The square of the splitting of the eigenvalues, $\Delta E_{ZZ} (\vec{k})=\left[E_{+}(\vec{k})-E_{+}(\vec{k})\right]/2$, can then be expressed as
\begin{equation}
\label{9a}
\left(\Delta E_{ZZ}(\vec{k})\right)^{2}= \left(\lambda_{1}k_{y} + \lambda_{3} k^{3}_{y} \right)^{2}.
\end{equation}
  
From Eq. (\ref{8}), it is revealed that $\hat{H}_{\textbf{SOC}}^{ZZ}$ is characterized only by one component of the wave vector, $k_{y}$. Therefore, $\hat{H}_{\textbf{SOC}}^{ZZ}$ preserves an ideal 1D Rashba SOC, akin to that observed in various VLD-based 2D TMDC systems \cite{Li2019, Adhib, Absor2024}. In addition, $\hat{H}_{\textbf{SOC}}$ is mainly determined by terms involving $\sigma_{x}$ and $\sigma_{z}$, indicating that the spin polarization is collinear but tilted along the $x-z$ plane in the FBZ. This analysis aligns well with the DFT results in our spin-resolved projected bands shown in Fig. 5(b) as well as the spin textures of the spin-split defect states presented in Figs. 5(c)-(d). Furthermore, as shown in Eq. (\ref{8}), both the $\sigma_{x}$ and $\sigma_{z}$ components of the spin vector remain conserved even at higher-order $\vec{k}$ terms, maintaining the unidirectional canted spin polarization at larger wave vectors $\vec{k}$ within the FBZ.  

On the contrary, only $k_{x}$ component in Eq. (\ref{7}) remains in the armchair-VLD systems since the spin-split defect states are confined and localized in the extended defect line along the $\Gamma-X$ ($k_{x}$) direction in the FBZ; see Fig. S1(c) of the Supplementary Materials \cite{Supplementary}. Thus, Eq. (\ref{7}) can be reformulated as
\begin{equation}
\label{10}
H_{\textbf{SOC}}^{AC}(\vec{k})= \alpha_{1}k_{x}\sigma_{y}+\alpha_{3}k^{3}_{x}\sigma_{y}+..+\alpha_{n}k^{n}_{x}\sigma_{y}
\end{equation}
This results in energy solution up to the third order term in $\vec{k}$,
\begin{equation}
\label{11}
E_{\pm}^{AC}(\vec{k})=E_{0}(\vec{k}) \pm \alpha_{1}k_{x} \pm \alpha_{3} k^{3}_{x},
\end{equation}
where $\alpha_{1}$ and $\alpha_{3}$ are the first and third order SOC strength parameters, respectively, along the $\Gamma-X$ ($k_{x}$) line. Therefore, the square of the splitting, $\left(\Delta E_{AC} (\vec{k})\right)^{2}$, can be written as
\begin{equation}
\label{11a}
\left(\Delta E_{AC}(\vec{k})\right)^{2}= \left(\alpha_{1}k_{x} + \alpha_{3} k^{3}_{x} \right)^{2}.
\end{equation}

Similar to the zigzag-VLD systems, $\hat{H}_{\textbf{SOC}}^{AC}$ of the armchair-VLD systems in Eq. (\ref{10}) also displays a perfect 1D Rashba SOC due to the preservation of only the $k_{x}$ component of the wave vector. However, as demonstrated in Eq. (\ref{10}), we clearly see that only $\sigma_{y}$ term characterizing the $\hat{H}_{\textbf{SOC}}$ of the armchair-VLD systems. This implies that the spin polarization becomes fully unidirectional oriented along the $k_{y}$-direction in the FBZ, in agreement with calculated spin-resolved projected bands of the spin-split defect states shown in Figs. S5(d)-(f) in the Supplementary Materials \cite{Supplementary}, respectively.

\begin{figure*}
	\centering
		\includegraphics[width=0.85\textwidth]{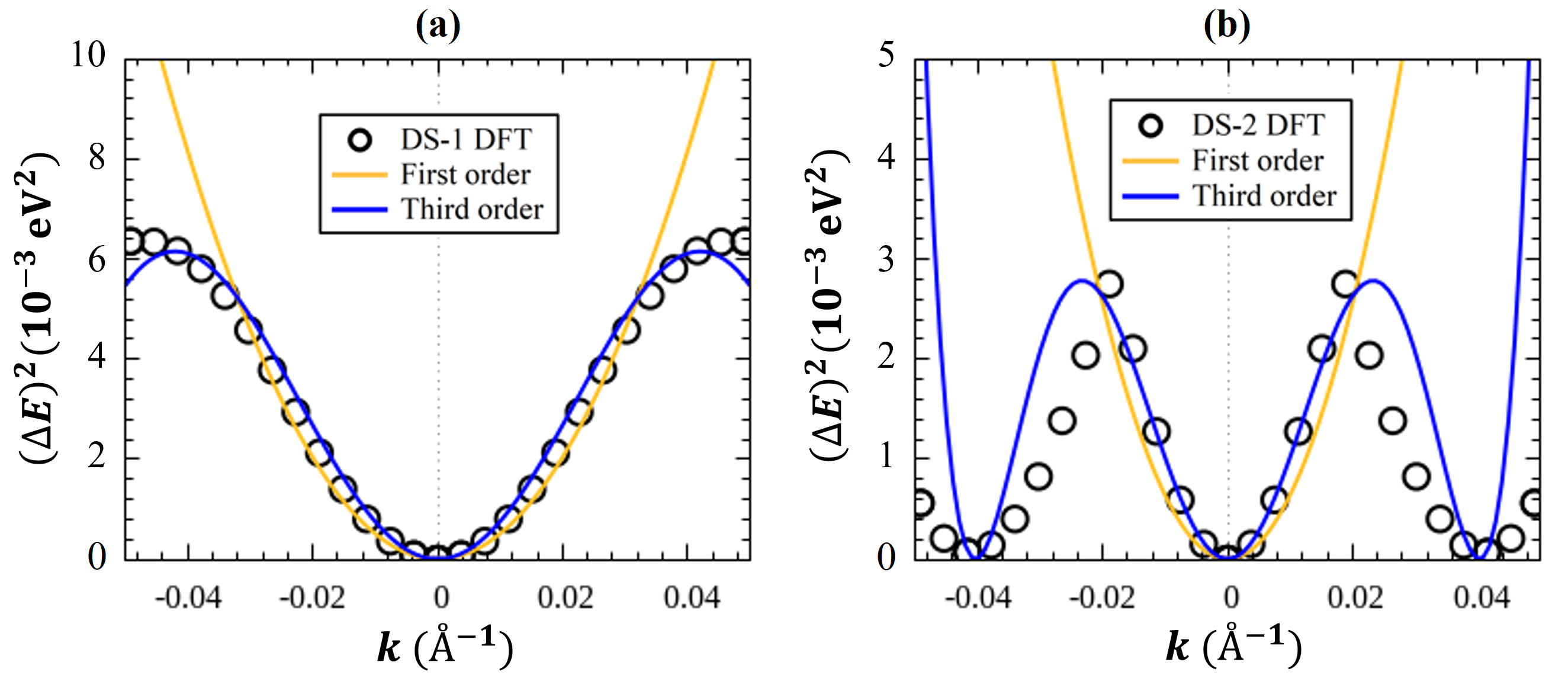}
	\caption{Square of the calculated splitting $\Delta E(\vec{k})=\left[E_{+}(\vec{k})-E_{+}(\vec{k})\right]/2$ of the defect states: (a) DS-1 and (b) DS-2, calculated along the $\Gamma-Y$ ($k_{y}$) are shown. Circle lines represent the DFT data, while solid lines correspond to the first-order (yellow) and third-order (blue) terms in $\vec{k}$ of the spin splitting, as obtained from the $\vec{k}\cdot\vec{p}$ Hamiltonian model in Eq. (\ref{9}), as described in the text.}
	\label{figure:Figure6}
\end{figure*}

\begin{table}[ht!]
\caption{The SOC strength parameters, $\lambda_{1}$ (eV\AA) and $\lambda_{3}$ (eV\AA$^{3}$) for the zigzag-VLD systems and $\alpha_{1}$ (eV\AA) and $\alpha_{3}$ (eV\AA$^{3}$) for armchair-VLD systems, are shown. The parameters were obtained by fitting the energy dispersion models in Eq. (\ref{9}) and Eq. (\ref{11}) to the spin-splitting of the DFT bands in the spin-split defect states (DS-1, DS-2). } 
\centering 
\begin{tabular}{ccc ccc ccc ccc} 
\hline\hline 
 1D defective systems &&& $\lambda_{1}$ ($\alpha_{1}$) [eV\AA]   &&& $\lambda_{3}$ ($\alpha_{3}$) [eV\AA$^{3}$]  &&& Reference    \\ 
\hline 
 $V_{\texttt{Si}}^{ZZ}$    &&& 2.52 (DS-1); 2.32 (DS-2)              &&& -508.34 (DS-1); -1105.8 (DS-2) &&&          This work   \\
 $V_{\texttt{Bi}}^{ZZ}$    &&& 2.60 (DS-1); 1.78 (DS-2)              &&& -146.6 (DS-1); -365.28 (DS-2)  &&&           This work    \\      
 $V_{\texttt{BiSi}}^{ZZ}$  &&& 3.28 (DS-1); 3.19 (DS-2)              &&& -110.56 (DS-1); -678.19 (DS-2) &&&           This work     \\
 $V_{\texttt{Si}}^{AC}$    &&& 2.93 (DS-1); 0.77 (DS-2)              &&& -255.28 (DS-1); -97.84 (DS-2)  &&&           This work   \\
 $V_{\texttt{Bi}}^{AC}$    &&& 2.25 (DS-1); 1.10 (DS-2)              &&& -48.4 (DS-1); -6.94 (DS-2)     &&&          This work    \\                         
 $V_{\texttt{SiBi}}^{AC}$  &&& 2.26 (DS-1); 3.08 (DS-2)              &&& -163.74 (DS-1); -317.19 (DS-2) &&&           This work     \\
 VLD on WTe$_{2}$ ML       &&&  0.09 - 3.64           &&&  (-17.31) - (-122.08)      &&&              Ref. \cite{Absor2024}    \\
 VLD on PtSe$_{2}$ ML      &&&  0.20 - 1.14           &&&                            &&&              Ref. \cite{Adhib}    \\ 
 1D Pb/Si(100)             &&&  1.11 - 1.14           &&&                            &&&                  Ref. \cite{Mihalyuk}    \\
 1D Pt/Si(110)             &&&  1.36                  &&&                            &&&                Ref. \cite{Park}    \\
 Bi-adsorbed 1D In chains  &&&  2.1 - 3.5             &&&                            &&&                 Ref. \cite{Tanaka}    \\
\hline\hline 
\end{tabular}
\label{table:Table 3} 
\end{table}

As a quantitative analysis, we further evaluate the SOC strength parameters by fitting the square of the splitting, $\left(\Delta E\right)^{2}$, in Eqs. (\ref{9a}) and (\ref{11a}) to the DFT bands in the spin-split defect states. In Figs. 6(a)-(b), the squared energy difference, $\left(\Delta E\right)^{2}$, is plotted along the $\Gamma-Y$ direction for the spin-split defect states (DS-1 and DS-2) of the $V_{\texttt{Si}}^{ZZ}$ system, alongside various fitting functions related to Eq. (\ref{9a}). A parabolic fit (yellow lines), $\left(\lambda_{1}k_{y}\right)^{2}$, with $\lambda_{1}=2.88$ eV\AA\ (DS-1) and $\lambda_{1}=2.31$ eV\AA\ (DS-2), matches the curve well only for $k<0.035$ \AA$^{-1}$ (DS-1) and $k<0.02$ \AA$^{-1}$ (DS-2). Up to $k\approx 0.05$ \AA$^{-1}$ (DS-1) and $k\approx 0.04$ \AA$^{-1}$ (DS-2), the curves remain consistent, but a third-order contribution becomes necessary for an accurate fit; see blue line in Figs. 6(a)-(b). Using the fitting function from Eq. (\ref{9a}), we determine $\lambda_{1}=2.52$ eV\AA\ (DS-1), $\lambda_{1}=2.32$ eV\AA\ (DS-2), $\lambda_{3}=-508.34$ eV\AA$^{3}$ (DS-1), and $\lambda_{3}=-1105.8$ eV\AA$^{3}$ (DS-1). Notably, the first-order $\lambda_{1}$ values for DS-1 and DS-2 align closely with those obtained from the third-order term, validating the accuracy of our fitting process. Additionally, the large magnitudes of the third-order SOC parameters ($\lambda_{3}$) highlight the critical role of the third-order term in Eq. (\ref{9}) for characterizing the spin-splitting behavior of the bands.

We then summarized the calculated results of the SOC strength parameters for all VLD systems in Table V and compare these results with a few selected 1D Rashba materials from previously reported data. In particular, the magnitude of the linear terms ($\alpha_{1}$, $\lambda_{1}$) of the SOC parameters for the defect states of the $V_{\texttt{Si}}^{ZZ}$ system [2.52 eV\AA\ (DS-1); 2.32 eV\AA\ (DS-2)] are much larger than that observed on the VLD-based PtSe$_{2}$ ML (0.20 eV\AA\ - 1.14 eV\AA) \cite{Adhib}, 1D Pb/Si(100) (1.11 eV\AA\ - 1.14 eV\AA) \cite{Mihalyuk}, and 1D Pt/Si(110) (1.36 eV\AA) \cite{Park}, and are comparable with that observed on the VLD-based WTe$_{2}$ (0.09 eV\AA\ - 3.64 eV\AA) \cite{Absor2024} and Bi-adsorbed 1D In chains (2.1 eV\AA\ - 3.5 eV\AA) \cite{Tanaka}. Remarkably, the associated SOC parameters found in the present VLD systems are sufficient to support the room temperature spintronics functionality \cite{Yaji2010}.

Finally, we explore the potential of utilizing VLD-engineered Si$_{2}$Bi$_{2}$ ML in spintronic devices. We propose that our VLD systems could serve as ideal 1D-Rashba channels in spin-field effect transistor (SFET) devices \cite{Chuang2015}. Electron spins could precess due to the Rashba effect, with the precession angle given by $\theta=2\alpha_{R}m^{*}L/\hbar^{2}$ (see Supplementary materials for detailed derivations \cite{Supplementary}), where $L$ is the distance between the source and drain, representing the length of the spin channel. By adjusting the gate voltage, we can tune the Rashba parameters $\alpha_{R}$, thus controlling the precession angle $\theta$. Owing to the large $\alpha_{R}$ in the VLD-based Si$_{2}$Bi$_{2}$ ML, our designed SFET should have a short channel length. For example, in the case of the $V_{\texttt{Si}}^{ZZ}$, we find that $\alpha_{R}=2.52$ eV\AA\ for DS-1 defect states, corresponding to a channel length of 2.5 nm. Indeed, this channel length is significantly smaller than the typical values of conventional SFETs ($\approx 2-5$ $\mu$m) operating at room temperature \cite{Takase2017, Chuang2015, Trier2020}. This short channel length favors the preservation of spin coherence and can be integrated into nanodevices with higher density. Moreover, since our VLD systems exhibit perfectly collinear spin polarization, they enable the generation of spin-polarized currents without dissipation, thereby enhancing SFET device performance.

\section{CONCLUSION} 

In conclusion, by combining first-principles DFT calculations with group theory analysis, we demonstrated that applying the VLDs to the Si$_{2}$Bi$_{2}$ ML results in perfect 1D Rashba states. We showed that VLDs introduce 1D confined defect states near the Fermi level, which are strongly localized along the defect line. These defect states exhibit significant 1D Rashba spin splitting, primarily due to strong $p-p$ coupling orbitals between Si and Bi atoms near the defect sites. Furthermore, these spin-split defect states exhibit unidirectional spin polarization in momentum $\vec{k}$-space, oriented perpendicularly to the VLD direction. Using $\vec{k}\cdot\vec{p}$ perturbation theory and symmetry analysis, we demonstrated that the 1D spin-split Rashba states with collinear spin polarization are enforced by the VLDs lowering the symmetry into the $C_{s}$ point group, which retains the $M_{xz}$ mirror symmetry and the 1D nature of the VLDs. These 1D Rashba states protect carriers against spin decoherence and support an exceptionally long spin lifetime \cite{Dyakonov, Schliemann, J_Schliemann}, making them promising for the development of efficient spintronic devices.

We highlight that our proposed method for inducing 1D Rashba states with significant spin splitting using the VLD is not confined to the Si$_{2}$Bi$_{2}$ ML. It can also be applied to other 2D group IV-V compounds, such as Si$_{2}X_{2}$ MLs ($X$ = P, As, Sb) and their Janus structures \cite{Ozdamar, Bafekry, Lee2020, Lukmantoro, Absor_Arif}, as well as In$_{2}$Se$_{2}$ \cite{Xiao2017, WangRSC, Wang2018} and Si$_{2}$P$_{2}$ \cite{Ha2024, Chen2015} MLs, which share similar structural symmetry and electronic properties. Recently, manipulation of the electronic properties of these particular materials by introducing defect was reported \cite{Lukmantoro, Xiao2017, WangRSC, Wang2018, Ha2024, Chen2015}. Therefore, it is expected that our predictions will stimulate further theoretical and experimental efforts in the exploration of the spin splitting properties of the 2D group IV-V compounds, expanding the range of 2D materials for future spintronic applications.

\begin{acknowledgments}

This work was supported by PD Research Grants (No.1709/UN1/DITLIT/Dit-Lit/PT.01.03/2022) funded by KEMDIKBUD-DIKTI, Republic of Indonesia. A.L. thanks the Indonesian Endowment Fund for Education (LPDP) Indonesia for financial support through the LPDP scholarship program. A.L. and E.S. thank BRIN for support through the research assistant (RA) program. The computation in this research was performed using the computer facilities at Gadjah Mada University and BRIN–HPC facilities. 

\end{acknowledgments}

\bibliography{Reference1}


\end{document}